\documentclass[letterpaper,twocolumn,english,prl,superscriptaddress,aps,longbibliography]{revtex4-2}
\usepackage{amsmath}
\usepackage{newtxmath}
\usepackage[T1]{fontenc}
\usepackage[latin9]{inputenc}
\usepackage{color}
\usepackage{babel}
\usepackage{mathdots}
\usepackage{stackrel}
\usepackage{graphicx}
\usepackage[pdfusetitle,
 bookmarks=true,bookmarksnumbered=false,bookmarksopen=false,
 breaklinks=false,pdfborder={0 0 1},backref=false,colorlinks=true]
 {hyperref}

\makeatletter

\usepackage{babel}
\usepackage{xcolor}
\usepackage{tikz}
\tikzset{every picture/.style={line width=0.75pt}}

\makeatother

\begin{document}
\title{Emergent non-Hermitian conservation laws at exceptional points}
\author{Zuo Wang}
\affiliation{Institute for Theoretical Physics, School of Physics, South China
Normal University, Guangzhou 510006, China}
\affiliation{Key Laboratory of Atomic and Subatomic Structure and Quantum Control
(Ministry of Education), Guangdong Basic Research Center of Excellence
for Structure and Fundamental Interactions of Matter, School of Physics,
South China Normal University, Guangzhou 510006, China}
\affiliation{Guangdong Provincial Key Laboratory of Quantum Engineering and Quantum
Materials, Guangdong-Hong Kong Joint Laboratory of Quantum Matter,
South China Normal University, Guangzhou 510006, China}
\author{Liang He}
\email{liang.he@scnu.edu.cn}

\affiliation{Institute for Theoretical Physics, School of Physics, South China
Normal University, Guangzhou 510006, China}
\affiliation{Key Laboratory of Atomic and Subatomic Structure and Quantum Control
(Ministry of Education), Guangdong Basic Research Center of Excellence
for Structure and Fundamental Interactions of Matter, School of Physics,
South China Normal University, Guangzhou 510006, China}
\affiliation{Guangdong Provincial Key Laboratory of Quantum Engineering and Quantum
Materials, Guangdong-Hong Kong Joint Laboratory of Quantum Matter,
South China Normal University, Guangzhou 510006, China}
\begin{abstract}
Non-Hermitian systems can manifest rich static and dynamical properties
at their exceptional points (EPs). Here, we identify yet another class
of distinct phenomena that is hinged on EPs, namely, the emergence
of a series of non-Hermitian conservation laws. We demonstrate these
distinct phenomena concretely in the non-Hermitian Heisenberg chain
and formulate a general theory for identifying these emergent non-Hermitian
conservation laws at EPs. By establishing a one-to-one correspondence
between the constants of motion (COMs) at EPs and those in corresponding
auxiliary Hermitian systems, we trace their physical origin back to
the presence of emergent symmetries in the auxiliary systems. Concrete
simulations on quantum circuits show that these emergent conserved
dynamics can be readily observed in current digital quantum computing
systems. 
\end{abstract}
\maketitle
\emph{Introduction}. Non-Hermitian Hamiltonians have recently been
found to exhibit remarkable applicability in a wide range of dynamical
systems. At the classical level, non-Hermitian Hamiltonians can capture
the distinct physics in complex systems with non-reciprocal interactions
ranging from active matters \cite{Uchida2010PRL,Nagy2010nature,Ivlev2015PRX,Lavergne2019science,Fruchart2021nature}
to metamaterials \cite{Scheibner2020natphys,Helbig2020natphys,Yang2024RMP}.
At the quantum level, they arise as effective descriptions of open
quantum systems \cite{ashida2020AP,Shibata2019PRB,wu2019science,xiao2020natphys,Liang2022PRL}
and continuously monitored quantum systems \cite{Fuji2020PRB,Jian2021PRB,Gopalakrishnan_Gullans_2021PRL,Turkeshi2021PRB,Buchhold2021PRX,Muller2022PRL},
providing crucial insight into the interplay between the quantum systems
and the environment. Even more remarkably, they can even describe
the physics associated to large-scale gravitational systems, ranging
from non-unitary photon absorption in black holes \cite{Svidzinsky2023PRD},
over Euclidean wormholes \cite{garcia2021PRD}, to emergent gravity
\cite{liu2023arXiv}.

One of the most intriguing characteristics of non-Hermitian systems
is the potential presence of exceptional points, which can give rise
to a plethora of distinct phenomena, including non-trivial topological
properties \cite{Bergholtz2021RMP,Ding2022NRP,Hu2022PRR,Lai_Nori_PRL_2024},
enhanced sensing \cite{Chen2017nature,hodaei2017nature,Xiao2019PRL,Zhang2019PRL},
dimension reduction \cite{chen2020nature}, and unconventional critical
phenomena \cite{Ashida2017natcomm,Hanai2020PRR,Fruchart2021nature,liu_Gong_Lee_2024arxiv}.
Quite often, EPs in quantum systems signify phase transitions due
to gap closing, as identified in excited-state quantum phase transitions
\cite{vsindelka2017PRA}, dynamical phase transitions \cite{eleuch2016PRA},
and measurement-induced entanglement transitions \cite{Gopalakrishnan_Gullans_2021PRL}.
Conventionally, gap closing indicates the emergence of new symmetries
if compared to the gapped case, such as those associated with scale
invariance \cite{Sachdev2001Cambridge} and broader classes of emergent
symmetries that go beyond the description provided by group theory
\cite{Ji2020PRR,Kong2020PRR}. Associated with the emergence of new
symmetries, new conservation laws can also arise. In these regards,
fundamental questions arise concerning the existence and properties
of the conservation laws at EPs, which could influence strongly the
macroscopic universal dynamical properties of the systems at EPs via
``slow modes'' \cite{Hohenberg_RMP_1977}.

\begin{figure}
\includegraphics[height=1.65in]{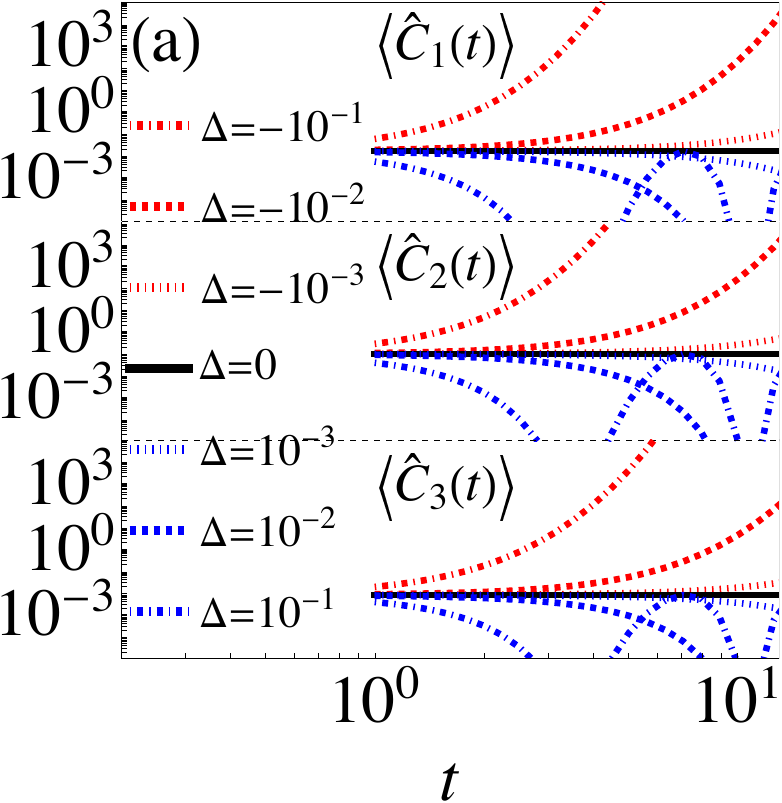}\includegraphics[height=1.65in]{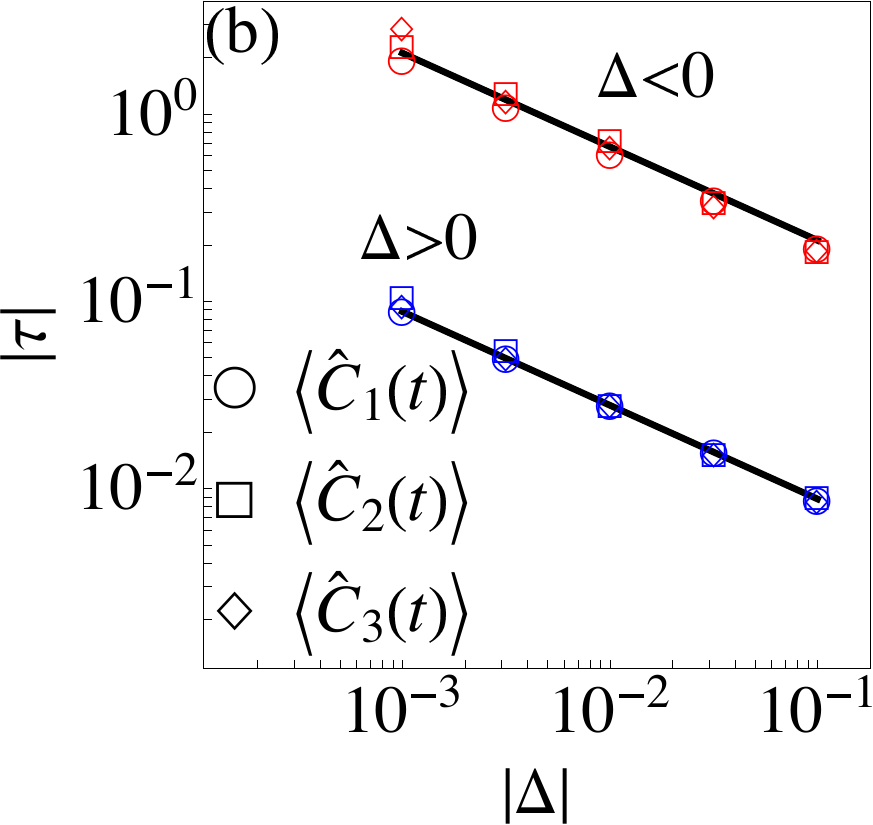}\caption{(a) Time evolution of $\langle\hat{C}_{1,\,2,\,3}\rangle(t)$ at
and away from EP for the non-Hermitian Heisenberg chain with $L=6$,
$J=1$, and the initial state $|\psi(t=0)\rangle=2^{-L}\sum_{\{\sigma_{j}^{z}\}}|\sigma_{1}^{z}\rangle\otimes\cdots\otimes|\sigma_{L}^{z}\rangle$.
$\langle\hat{C}_{1,\,2,\,3}\rangle(t)$ exhibit exponential growth
(oscillations) for $\Delta<0$ ($\Delta>0$). As $\Delta$ approaches
zero from either the positive or negative side, $\langle\hat{C}_{1,\,2,\,3}\rangle(t)$
changes slower and finally keeps at a constant with respect to time.
(b) Scaling behavior of the characteristic timescale $\tau$ with
respect to $|\Delta|$ in the proximity of the EP (the black solid
line $|\tau|\propto|\Delta|^{-1/2}$ is a guide for eyes). The characteristic
timescale $\tau$ is obtained by fitting $\langle\hat{C}_{1,\,2,\,3}\rangle(t)$
concerning $Ae^{t/\tau}$ and $B\cos(t/\tau+\theta)+C$ ($A$, $B$,
$C$, $\theta$ and $\tau$ are real fitting parameters), respectively
for $\Delta<0$ and $\Delta>0$. See text for more details.}\label{fig: C_1_2_3_time_evolution}
\end{figure}

In this paper, we address these questions in both a concrete non-Hermitian
Heisenberg chain and generic non-Hermitian quantum systems, and find
the following. (i) Emergence of a series of non-Hermitian conservation
laws at EPs. For the non-Hermitian Heisenberg chain, a series of non-Hermitian
conservation laws are identified at its EPs (see Fig.~\ref{fig: C_1_2_3_time_evolution}).
Furthermore, a general framework is established for identifying non-Hermitian
conservation laws that emerge at EPs of generic non-Hermitian quantum
systems by employing generalized eigenvectors {[}see Eq.~(\ref{eq: COM_general_EP}){]}.
(ii) The emergence of these non-Hermitian conservation laws can be
traced back to the presence of emergent symmetries in the auxiliary
Hermitian systems. Specifically, for the non-Hermitian Heisenberg
chain, its emergent non-Hermitian conservation laws can be traced
back to the presence of SU(2) symmetry in the auxiliary Hermitian
isotropic Heisenberg chain. (iii) These emergent conserved dynamics
can be observed in current digital quantum computing systems (see
Fig.~\ref{fig: IBMQC}).

\emph{Emergent non-Hermitian conservation laws at the EPs of a non-Hermitian
Heisenberg chain}. Let us first consider a non-Hermitian spin-$1/2$
Heisenberg chain described by the following Hamiltonian: 
\begin{equation}
\hat{H}_{\mathrm{NHS}}=\hat{H}_{\mathrm{XXX}}+g\sum_{j=1}^{L}\hat{\sigma}_{j}^{x}+ig(1-\Delta)\sum_{j=1}^{L}\hat{\sigma}_{j}^{y},\label{eq: Heisenberg_nhh}
\end{equation}
where $\hat{H}_{\text{XXX}}\equiv J\sum_{j=1}^{L-1}\sum_{a=x,y,z}\hat{\sigma}_{j}^{a}\hat{\sigma}_{j+1}^{a}$
is the Hamiltonian for the XXX (isotropic) Heisenberg chain with $J$
being the spin interaction strength and $(\hat{\sigma}^{x},\hat{\sigma}^{y},\hat{\sigma}^{z})$
being Pauli operators. The second term of $\hat{H}_{\mathrm{NHS}}$
is a transverse field term with $g$ being the field strength. The
third term is a non-Hermitian term with its strength parameterized
as $g(1-\Delta)$ with $\Delta$ being a real number. This type of
non-Hermitian Hamiltonian usually arises in quantum systems subjected
to continuous measurements and post-selections \cite{Biella2021Quantum,Gopalakrishnan_Gullans_2021PRL,Wang2023arXiv}.
Here, the non-Hermitian term in $\hat{H}_{\mathrm{NHS}}$ encapsulates
the physical effects of weak measurements along the \ensuremath{y}-direction
and subsequent post-selections \cite{Wang2023arXiv}. In the following,
we use the units with $g=1$ and $\hbar=1$.

From the form of $\hat{H}_{\mathrm{NHS}}$, we note that $\Delta<0$
and $\Delta>0$ signify phases with parity-time symmetry breaking
and preserving, respectively, and at the transition point $\Delta=0$,
EPs of different orders emerge in $\hat{H}_{\mathrm{NHS}}$ \cite{Wang2023arXiv},
accompanied by coalesce of eigenvectors and closing of gaps between
eigenenergies. Frequently, gap closing could indicate the emergence
of higher symmetries \cite{Sachdev2001Cambridge,Ji2020PRR,Kong2020PRR}
and associated conservation laws. In the following, we focus on whether
non-Hermitian conservation laws could emerge in this scenario and
their properties.

Since for a generic non-Hermitian quantum system with completely real
spectrum described by the Hamiltonian $\hat{H}$, its conserved quantities
or non-Hermitian COMs, denoted as $\hat{O}$, should satisfy the condition
$[\hat{H}^{\dagger}\hat{O}-\hat{O}\hat{H}]=0$ to guarantee its average
value being a constant with respect to time, i.e., $\partial_{t}\langle\psi(t)|\hat{O}|\psi(t)\rangle=0$
($|\psi(t)\rangle$ is the state of the system at time $t$) \citep{Wang2022PRB, Bian2020PRR,ruzicka2021conserved}.
To probe the existence of conserved quantities, we enumerate Hermitian
operators for a small non-Hermitian Heisenberg chain (e.g., $L=3$),
and identify those that satisfy the conservation condition $[\hat{H}_{\text{NHS}}^{\dagger}\hat{O}-\hat{O}\hat{H}_{\text{NHS}}]=0$
at $\Delta=0$. We then generalize these operators to arbitrary chain
lengths $L$. Using this brute-force method, we indeed identify two
COMs, $\hat{C}_{1}$ and $\hat{C}_{2}$ with the explicit forms 
\begin{align}
 & \hat{C}_{1}=(\hat{P}_{\downarrow}){}^{\otimes L},\,\hat{C}_{2}=\frac{1}{L}\sum_{i=1}^{L}(\hat{P}_{\downarrow})^{\otimes(i-1)}\otimes\frac{\hat{\sigma}_{i}^{x}}{2}\otimes(\hat{P}_{\downarrow})^{\otimes(L-i)},\label{eq: Two_Concise_EOMs}
\end{align}
where $\hat{P}_{\downarrow}\equiv|\downarrow\rangle\langle\downarrow|$
is the local projection operator along $z$-direction.

The time evolution for the expectation values $\langle\hat{C}_{1,\,2}\rangle(t)\equiv\langle\psi(t)|\hat{C}_{1,\,2}|\psi(t)\rangle$
of the non-Hermitian Heisenberg chain with $L=6$, $J=1$, and the
initial state $|\psi(t=0)\rangle=2^{-L}\sum_{\{\sigma_{j}^{z}\}}|\sigma_{1}^{z}\rangle\otimes\cdots\otimes|\sigma_{L}^{z}\rangle$
is shown in Fig.~\ref{fig: C_1_2_3_time_evolution}(a). We can directly
see that indeed, as $\Delta$ approaches zero from either the positive
or negative side, $\langle\hat{C}_{1,\,2}\rangle(t)$ changes slower
and finally keeps at a constant with respect to time, signifying that
they emerge as COM at $\Delta=0$ (see the Supplemental Material \cite{Sup_Mat}
for a further demonstration of their robustness in a modified model
incorporating weak on-site noise). Moreover, to quantitatively analyze
the emergence of these COMs, we determine a characteristic timescale
$\tau$ by performing fits of a simple exponential function $Ae^{t/\tau}$
and an oscillatory function $B\cos(t/\tau+\theta)+C$ ($A$, $B$,
$C$, $\theta$ and $\tau$ are real fitting parameters) to the numerical
data of $\langle\hat{C}_{1,\,2}\rangle(t)$ for $\Delta<0$ and $\Delta>0$,
respectively. As shown in Fig.~\ref{fig: C_1_2_3_time_evolution}(b),
the emergence of these COMs coincides with the divergence of the characteristic
timescale $\tau$. Interestingly, we observe a power law scaling of
$\tau$ with respect $|\Delta|$, $\tau\propto|\Delta|^{-1/2}$, which
can be traced back to properties of the parity-time symmetry breaking
in this system \citep{Wang2023arXiv}.

\begin{figure}
\includegraphics[width=2.5in]{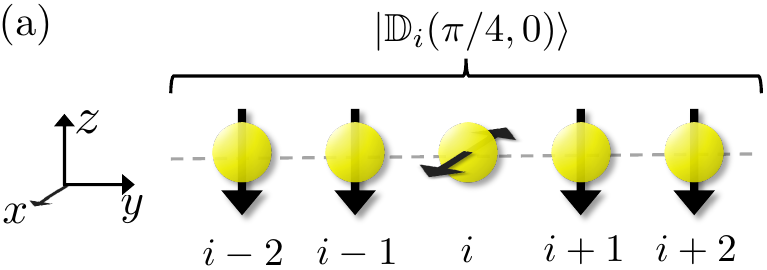} \includegraphics[width=3in]{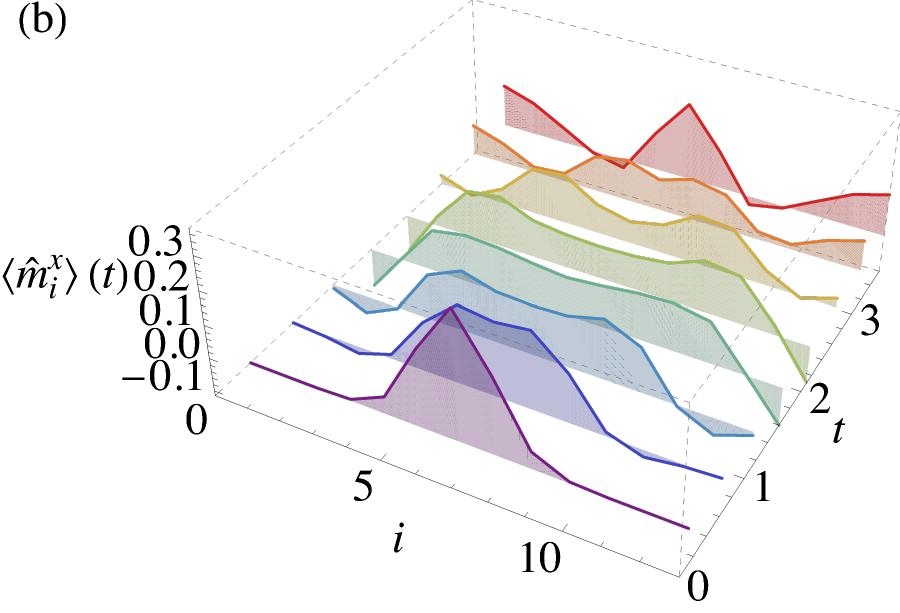}
\caption{(a) Schematic illustration of the completely polarized background
state with a single \textquotedblleft defect\textquotedblright{} on
site $i$, i.e., $|\mathbb{D}_{i}(\pi/4,0)\rangle$. (b) Time evolution
of the spatial density distribution $\langle\hat{m}_{i}^{x}\rangle(t)$
of the COM $\langle\hat{C}_{2}\rangle(t)$ at $\Delta=0$, with $L=13$,
$J=1$ and initial state being $2^{1/2}|\mathbb{P}\rangle+\sum_{i=1}^{L}2^{1/2}(\mathcal{N}\exp(-(i-7)^{2}/2)|\mathbb{D}_{i}(\pi/2,0)\rangle)$
($\mathcal{N}$ is a normalization factor). The time evolution of
the spatial distribution $\langle\hat{m}_{i}^{x}\rangle(t)$ is driven
by the Noether current $\hat{j}_{i}$ according to the continuity
equation (\ref{eq: Continuity_EQ}), whereas the expectation value
of its corresponding COM $\hat{C}_{2}=\sum_{i}\hat{m}_{i}^{x}$, remains
constant with respect to time ($\langle\hat{C}_{2}\rangle(t)=0.941$
in this case). See text for more details.}\label{fig:  DiagramConservation}
\end{figure}

\begin{figure}
\includegraphics[width=2.6in]{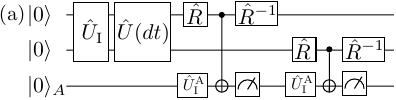}\\
 ~

\includegraphics[width=2.1in]{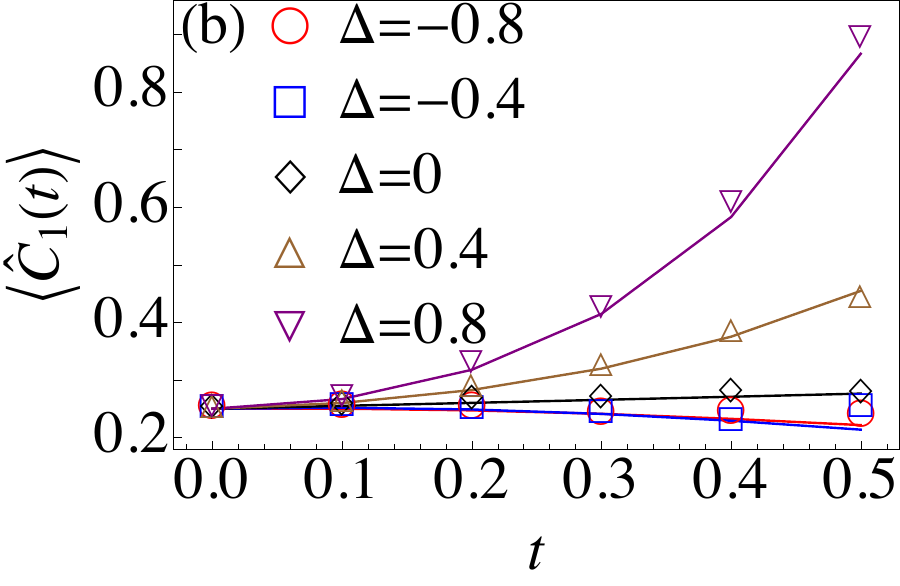}\caption{Quantum circuit simulation of $\langle\hat{C}_{1}(t)\rangle$ in
a $2$-site non-Hermitian Heisenberg chain. (a) Quantum circuit protocol
for simulating the dynamics in the first time step. (b) Simulated
dynamics of $\langle\hat{C}_{1}(t)\rangle$. The markers represent
the results obtained from $10^{6}$ trajectories by \emph{qasm\_simulator}
with the noise from \emph{ibm\_lagos} from IBM Qiskit \citep{cross2018ibm}.
The corresponding solid curves are the exact expectations obtained
by numerical calculations. All dynamics are simulated with parameters
$dt=0.1$, $J=1$. See text for more details.}\label{fig: IBMQC}
\end{figure}

The above COMs $\hat{C}_{1,2}$ also assume clear physical meanings.
For $\hat{C}_{1}$, we notice that it is simply a projection operator
to the completely polarized state, denoted as $|\mathbb{P}\rangle\equiv\otimes_{i=1}^{L}|\downarrow_{i}\rangle$.
The conservation of $\langle\hat{C}_{1}\rangle$ is nothing but the
conservation of probability of the system in the state $|\mathbb{P}\rangle$.
For $\hat{C}_{2}$, we can see that it is a sum over the local operator
$\hat{m}_{i}^{x}\equiv(\hat{P}_{\downarrow})^{\otimes(i-1)}\otimes(\hat{\sigma}_{i}^{x}/2)\otimes(\hat{P}_{\downarrow})^{\otimes(L-i)}$
, i.e., $\hat{C}_{2}=(1/L)\sum_{i=1}^{L}\hat{m}_{i}^{x}$. The expectation
value of $\hat{m}_{i}^{x}$ corresponds to the $x$-magnetization
of the completely polarized background state with a single ``defect''
on site $i$, i.e., $|\mathbb{D}_{i}(\theta,\phi)\rangle\equiv\otimes_{j=1}^{i-1}|\downarrow_{j}\rangle\otimes|d_{i}(\theta,\phi)\rangle\otimes_{k=i+1}^{L}|\downarrow_{k}\rangle$
with $\theta,\phi\in\text{Re}$ and $|d_{i}(\theta,\phi)\rangle=e^{i\phi}\cos\theta|\downarrow_{i}\rangle+\sin\theta|\uparrow_{i}\rangle$
being a generic state on site $i$ (see Fig.~\ref{fig:  DiagramConservation}(a)
for a schematic illustration). Therefore, the conservation of $\langle\hat{C}_{2}\rangle$
corresponds to the conservation of total $x$-magnetization in the
single ``defect'' subspace spanned by $\{|\mathbb{P}\rangle,|\mathbb{D}_{i}(\pi/2,0)\rangle|i=1,\dots L\}$,
and $\hat{m}_{i}^{x}$ is spatial density of this conserved quantity.
Interestingly, one can further construct a Noether current operator
$\hat{j}_{i}\equiv J[(\hat{P}_{\downarrow})^{\otimes i}\otimes\hat{\sigma}_{i+1}^{y}\otimes(\hat{P}_{\downarrow})^{\otimes(L-i-1)}-(\hat{P}_{\downarrow})^{\otimes(i-1)}\otimes\hat{\sigma}_{i}^{y}\otimes(\hat{P}_{\downarrow})^{\otimes(L-i)}]$
(with $i=1,\dots L-1$ and $\hat{j}_{i}=0$ for other $i$) that gives
rise to the continuity equation (discrete version) for $\hat{C}_{2}$,
i.e., 
\begin{equation}
\partial_{t}\langle\hat{m}_{i}^{x}\rangle(t)+\langle\hat{j}_{i}\rangle(t)-\langle\hat{j}_{i-1}\rangle(t)=0.\label{eq: Continuity_EQ}
\end{equation}
This equation governs the dynamics of the density distribution of
$\hat{C}_{2}$ as showcased in Fig.~\ref{fig:  DiagramConservation}(b)
for a non-Hermitian Heisenberg chain with $L=13$ at $\Delta=0$.

From the existence of $\hat{C}_{1,2}$, we see that non-Hermitian
COMs can indeed emerge at exceptional points. Furthermore, one expects
there may exist even more COMs besides $\hat{C}_{1,2}$. To find the
complete set of COMs, we now develop a systematic approach for constructing
COMs using the so-called generalized eigenvectors associated with
EPs \cite{bronson1991matrix}. Without loss of generality, let us
consider a generic non-Hermitian Hamiltonian $\hat{H}$ that assumes
an EP with $N$ eigenstates coalesce to a single eigenstate with real
energy $E$. The $N$ generalized eigenvectors for the conjugated
Hamiltonian, denoted as $|V_{1}\rangle,\dots,|V_{N}\rangle$, can
be constructed according to $\hat{H}^{\dagger}|V_{n}\rangle=E|V_{n}\rangle+|V_{n-1}\rangle$,
where $1\leq n\leq N$ and $|V_{0}\rangle\equiv0$ \cite{bronson1991matrix}.
Using these generalized eigenvectors, we can decompose any operators
into the linear combinations of a set of operators $\{\hat{V}_{i,j}\equiv|V_{i}\rangle\langle V_{j}||i,j=1,2,\dots,N\}$
constructed by them. It can be straightforwardly verified (see Supplemental
Material \cite{Sup_Mat}) that any COM associated with this EP must
be a linear combination of the following $N$ independent COMs, 
\begin{equation}
\hat{C}_{n}=\sum_{j=1}^{n}\hat{V}_{j,n-j+1},~1\leq n\leq N.\label{eq: COM_general_EP}
\end{equation}
For the non-Hermitian Heisenberg chain (\ref{eq: Heisenberg_nhh}),
let us focus on the EP of order $L+1$ for $\hat{H}_{\mathrm{NHS}}(\Delta=0)$
\citep{Wang2023arXiv}. We find that in this case, the generalized
eigenvectors are proportional to the $(L+1)$-fold degenerate eigenstates
of the Heisenberg XXX model $\hat{H}_{\mathrm{XXX}}$, i.e., 
\begin{equation}
|V_{n}\rangle=c_{n}\left(\sum_{i=1}^{L}\frac{\hat{\sigma}_{i}^{+}}{2}\right)^{n-1}\stackrel[j=1]{L}{\otimes}|\downarrow_{j}\rangle,\label{eq: GE_Heisenberg}
\end{equation}
where $n=1,2,\cdots,L+1$, $c_{n}\equiv2^{1-n}(L+1-n)!/L!/(n-1)!$,
$\hat{\sigma}^{\pm}\equiv\hat{\sigma}^{x}\pm i\hat{\sigma}^{y}$.
Expressing the COMs in (\ref{eq: COM_general_EP}) in terms of the
products of Pauli operators, we observe that the first two independent
COMs exactly take the form of two COMs we find in (\ref{eq: Two_Concise_EOMs}),
i.e., $\hat{C}_{1}=\hat{V}_{1,1}=(\hat{P}_{\downarrow}){}^{\otimes L}$
and $\hat{C}_{2}=\sum_{j=1}^{2}\hat{V}_{j,3-j}=(1/L)\sum_{i=1}^{L}\hat{m}_{i}^{x}$.
The third COM in (\ref{eq: COM_general_EP}) is given by $\hat{C}_{3}=\sum_{j=1}^{3}\hat{V}_{j,4-j}$,
whose explicit form reads 
\begin{align}
\hat{C}_{3}= & \sum_{\sigma=\pm}\sum_{i,j=1}^{L}\frac{1}{8L^{2}}\left(\frac{L}{L-1}\hat{m}_{i,j}^{\sigma_{i},\sigma_{j}}+\hat{m}_{i,j}^{\sigma_{i},\bar{\sigma}_{j}}\right)-\frac{1}{4L}\hat{C}_{1}\label{eq: COM3_Heisenberg}
\end{align}
with $\hat{m}_{i_{1},i_{2}}^{\sigma_{i_{1}},\sigma_{i_{2}}^{\prime}}\equiv\hat{P}_{\downarrow}^{\otimes(i_{p(1)}-1)}\otimes(\hat{\sigma}_{i_{p(1)}}^{\sigma_{i_{p(1)}}}/2)\otimes\hat{P}_{\downarrow}^{\otimes(i_{p(2)}-i_{p(1)}-1)}\otimes(\hat{\sigma}_{i_{p(2)}}^{\sigma_{i_{p(2)}}^{\prime}}/2)\otimes\hat{P}_{\downarrow}^{\otimes(L-i_{p(2)})}$
($\sigma,\sigma^{\prime}=\pm$ and $i_{p(k)}$ is the $k$th smallest
number in set $\{i_{1},i_{2}|i_{2}\neq i_{1}\}$), $\hat{m}_{i_{1},i_{2}=i_{1}}^{\sigma_{i_{1}},\sigma_{i_{2}}^{\prime}=\sigma_{i_{1}}}=0$,
and $\hat{m}_{i,i}^{\pm_{i},\mp_{i}}\equiv\hat{P}_{\downarrow}^{\otimes(i-1)}\otimes(\hat{\sigma}_{i}^{\ensuremath{\pm_{i}}}\hat{\sigma}_{i}^{\mp_{i}}/2)\otimes\hat{P}_{\downarrow}^{\otimes(L-i)}$.
From the explicit form of $\hat{C}_{3}$, we can see that it is a
sum over all the ``pair'' operators $\hat{m}_{i,j}^{\sigma_{i},\sigma_{j}^{\prime}}$,
the expectation value of which corresponds to the two-point correlation
for the magnetization of the completely polarized background state
with a pair of ``defects'' locating on site $i$ and $j$, respectively,
i.e., $|\mathbb{D}_{i}(\theta_{i},\phi_{i})\mathbb{D}_{j}(\theta_{j},\phi_{j})\rangle\equiv\otimes_{k=1}^{i-1}|\downarrow_{k}\rangle\otimes|d_{i}(\theta,\phi)\rangle\otimes_{k=i+1}^{j-1}|\downarrow_{k}\rangle\otimes|d_{j}(\theta,\phi)\rangle\otimes_{k=j+1}^{L}|\downarrow_{k}\rangle$
for $i\neq j$ and $|\mathbb{D}_{i}(\theta,\phi)\rangle$ for $i=j$.
Although the explicit expressions for COMs $\hat{C}_{n>3}$ become
lengthy, one is still able to obtain all remaining COMs related to
the $(L+1)$-fold degenerate subspace by straightforward calculations
according to the general form of $\hat{C}_{n}$ shown in Eq.~(\ref{eq: COM_general_EP}).

\emph{Origin of the COMs at EP.} Having shown that a series of non-Hermitian
COMs can emerge at EP, we shall now trace their physical origin. To
this end, we take an approach that employs an auxiliary \textit{Hermitian}
system. More specifically, for the non-Hermitian Heisenberg spin chain
$\hat{H}_{\mathrm{NHS}}$ with $0<\Delta<2$, it is diagonalizable
with its spectrum being completely real. Therefore, according to a
generic albeit abstract construction in terms of the left and right
eigenstates of $\hat{H}_{\mathrm{NHS}}$ \cite{mostafazadeh2002JMP},
there exists a similarity transformation $\hat{\mathcal{S}}$, by
which $\hat{H}_{\mathrm{NHS}}$ can be transformed to an auxiliary
Hermitian Hamiltonian, denoted as $\hat{H}_{\mathrm{AHS}}$, that
assumes the same spectrum, i.e., $\hat{H}_{\mathrm{AHS}}=\hat{\mathcal{S}}\hat{H}_{\mathrm{NHS}}\hat{\mathcal{S}}^{-1}$.
Despite the abstract construction of $\hat{\mathcal{S}}$ does not
give its form concretely, here we can find a concrete concise form
of $\hat{\mathcal{S}}$ expressed in terms of the local spin operators
of the system, i.e., $\hat{\mathcal{S}}=[\exp(\hat{\sigma}^{z}/2\ln\sqrt{\Delta/(2-\Delta)})]^{\otimes L}$.
This result can be obtained by first examining the single-spin case
and then constructing a general ansatz (see Supplemental Material
\cite{Sup_Mat} for derivation details). Under this similar transformation,
the auxiliary Hermitian Hamiltonian also assumes a concise form of
a transverse Hermitian Heisenberg model, i.e.,

\begin{equation}
\hat{H}_{\mathrm{AHS}}=\hat{\mathcal{S}}\hat{H}_{\mathrm{NHS}}\hat{\mathcal{S}}^{-1}=\hat{H}_{\mathrm{XXX}}+\sum_{j=1}^{L}\sqrt{\Delta(2-\Delta)}\hat{\sigma}_{j}^{x}.\label{eq: similar_Hamiltonian_Heisenberg}
\end{equation}

From (\ref{eq: similar_Hamiltonian_Heisenberg}), we see that the
auxiliary Hermitian system $\hat{H}_{\mathrm{AHS}}$ assumes the rotation
symmetry with respect to the $x$-direction for finite $\Delta$,
and approaches $\hat{H}_{\mathrm{XXX}}$ which assumes a higher SU(2)
symmetry as $\Delta$ goes to zero. Moreover, as pointed out in previous
investigations \cite{Wang2022PRB}, each COM present in the auxiliary
Hermitian system, denoted as $\hat{C}^{A}$ which satisfies $[\hat{C}^{A},\hat{H}_{\mathrm{AHS}}]=0$,
generally maps to a non-Hermitian COM, denoted as $\hat{C}$, in the
corresponding non-Hermitian system via the same similarity transformation,
i.e., $\hat{C}=\hat{\mathcal{S}}^{\dagger}\hat{C}^{A}\hat{\mathcal{S}}$.
In these regards, one naturally expects that the emergent conservation
laws at EP can be traced back to the presence of higher symmetry in
the auxiliary Hermitian system $\hat{H}_{\mathrm{AHS}}$ when $\Delta$
goes to zero.

Indeed, we find that for each $\hat{C}_{n}$ at EP, there exists a
corresponding $\hat{C}_{n}^{A}$ for $\left.\hat{H}_{\mathrm{AHS}}\right|_{\Delta=0}$
with the higher $\mathrm{SU(2)}$ symmetry, i.e., 
\begin{align}
\hat{C}_{n} & =\lim_{\Delta\rightarrow0^{+}}\hat{\mathcal{S}}^{\dagger}\hat{C}_{n}^{A}\hat{\mathcal{S}},\,\label{eq: COM_AHS_NHS correspondance}
\end{align}
with $\hat{C}_{n}^{A}=[\Delta/(2-\Delta)]^{(L-n+1)/2}\hat{C}_{n}$.
Interestingly, we notice that $\hat{C}_{n}^{A}$ assumes essentially
the same form as $\hat{C}_{n}$ up to a pre-factor. This originates
from the coincidence that the generalized eigenvectors in (\ref{eq: GE_Heisenberg})
are also eigenvectors of $\hat{\mathcal{S}}$, i.e., $\hat{\mathcal{S}}|V_{n}\rangle=[\Delta/(2-\Delta)]^{(n-1)-L/2}|V_{n}\rangle$,
so that $\hat{\mathcal{S}}^{\dagger}\hat{C}_{n}\hat{\mathcal{S}}=[\Delta/(2-\Delta)]^{(-L+n-1)/2}\hat{C}_{n}$.
Moreover, we remark that although each non-Hermitian COM at EP can
be traced back to a COM for the auxiliary highly symmetric Hermitian
system $\left.\hat{H}_{\mathrm{AHS}}\right|_{\Delta=0}=\hat{H}_{\mathrm{XXX}}$,
the converse is not true: not every COM of $\left.\hat{H}_{\mathrm{AHS}}\right|_{\Delta=0}$
corresponds to a COM in the non-Hermitian system. For instance, any
operator that is proportional to the total magnetization along $z$-direction
$\sum_{i=1}^{L}\hat{\sigma}_{i}^{z}/2$ is naturally a COM of $\hat{H}_{\mathrm{XXX}}$.
However, it does not give rise to any well-defined non-Hermitian COM
via Eq.~(\ref{eq: COM_AHS_NHS correspondance}) due to the presence
of divergence that cannot be properly regularized \footnote{For instance, we consider the COM of $\hat{H}_{\mathrm{XXX}}$, $f(\Delta)\sum_{i=1}^{L}\hat{\sigma}_{i}^{z}/2$,
where $f(\Delta)$ takes a finite value as $\Delta\rightarrow0^{+}$.
Plugging $f(\Delta)\sum_{i=1}^{L}\hat{\sigma}_{i}^{z}/2$ into the
right-hand side of Eq.~(\ref{eq: COM_AHS_NHS correspondance}), we
find $\hat{C}_{n}=\lim_{\Delta\rightarrow0^{+}}\hat{\mathcal{S}}^{2}\left[f(\Delta)\sum_{i=1}^{L}\hat{\sigma}_{i}^{z}/2\right]$.
Noticing that the matrix representation of $\hat{\mathcal{S}}^{2}=\left[\left(\Delta/(2-\Delta)\right)^{\hat{\sigma}^{z}/2}\right]^{\otimes L}$
contains divergences of the form $\Delta^{-1/2}$, $\Delta^{-1}$,
$\Delta^{-3/2}$, etc., and these divergences cannot be removed by
multiplying a global value $f(\Delta)$, we find $\hat{C}_{n}$ cannot
be a physical observable in the finite-dimensional Hilbert space. }.

From the above discussion we have seen that for the non-Hermitian
Heisenberg chain, the COMs at EP can be traced back to COMs present
in the highly symmetric auxiliary Hermitian system. In fact, this
is generally true for any non-Hermitian system at EP, as we shall
now briefly show (see \cite{Sup_Mat} for more details). For a generic
non-Hermitian system with EP, one can always find a suitable representation
under which the matrix representation of its Hamiltonian assumes a
``deformed'' Jordan normal form, i.e., $\mathbf{H}=\oplus_{i=1}^{i_{\max}}(E_{i}\mathbf{I}_{N_{i}}+\mathbf{S}_{N_{i}}^{x}+i\mathbf{S}_{N_{i}}^{y})$
\cite{bronson1991matrix,Sup_Mat}. Here, $\mathbf{I}_{N_{i}}$ is
a $N_{i}\times N_{i}$ identity matrix, $\mathbf{S}_{N_{i}}^{x(y)(z)}$
is the $N_{i}\times N_{i}$ spin matrices for spin-$(N_{i}-1)/2$
operators along the $x(y)(z)$-direction in the $z$-basis (the conventions
$\hbar=1$ and $\mathbf{S}_{N_{i}=1}^{x}=\mathbf{S}_{N_{i}=1}^{y}=\mathbf{S}_{N_{i}=1}^{z}=0$
are adopted), and $E_{i}$ is the corresponding real eigenvalue for
the $i$th block of $\mathbf{H}$. Now, we can introduce a ``perturbed''
matrix $\mathbf{H}(\Delta)=\stackrel[i=1]{i_{\max}}{\oplus}(E_{i}\mathbf{I}_{N_{i}}+\mathbf{S}_{N_{i}}^{x}+i(1-\Delta)\mathbf{S}_{N_{i}}^{y})$
with $\Delta$ being the strength of the ``perturbation'', and reproducing
$\mathbf{H}$ in the $\Delta\rightarrow0^{+}$ limit. Straightforwardly,
one can show that $\mathbf{H}(\Delta)$ can be mapped to a Hermitian
matrix $\mathbf{H}_{\mathrm{A}}(\Delta)\equiv\stackrel[i=1]{i_{\max}}{\oplus}(E_{i}\mathbf{I}_{N_{i}}+\sqrt{\Delta(2-\Delta)}\mathbf{S}_{N_{i}}^{x})$
via the similarity transformation $\mathbf{H}_{\mathrm{A}}(\Delta)=\mathbf{S}\mathbf{H}(\Delta)\mathbf{S}^{-1}$
with $\mathbf{S}=\stackrel[i=1]{i_{\max}}{\oplus}\exp(\mathbf{S}_{N_{i}}^{z}\ln\sqrt{\Delta/(2-\Delta)})$.
From the form of $\mathbf{H}_{\mathrm{A}}(\Delta)$, we notice that
in the limit $\Delta\rightarrow0^{+}$, it assumes a higher symmetry
demonstrated by the emergence of degeneracy within each block with
$N_{i}>1$. This gives rise to the same scenario as the non-Hermitian
Heisenberg chain. And indeed, for any COM of the non-Hermitian system
at EP described by $\mathbf{H}$, it can be traced back to the COM
of the auxiliary Hermitian system described by $\mathbf{H}_{\mathrm{A}}(\Delta=0)$
that assumes a higher symmetry (see \cite{Sup_Mat} for more details).
We further note that although $\mathbf{H}_{\mathrm{A}}(\Delta=0)$
can generally be constructed as shown above, expressing it in terms
of the system's elementary operators can still be challenging. However,
for systems whose Hamiltonians consist of $\mathrm{\text{SU}}(2)$-symmetric
operators and operators that generate the $\mathfrak{su}(2)$ Lie
algebra---such as the non-Hermitian Heisenberg chain $\hat{H}_{\mathrm{NHS}}$
considered here, the isotropic Heisenberg spin system in arbitrary
dimensions with local transverse fields and the Fermi-Hubbard model---one
can can indeed express $\mathbf{H}_{\mathrm{A}}(\Delta=0)$ in terms
of the elementary operators of the system by employing the general
form of the similarity transformation $\mathbf{S}$ introduced above
(see \cite{Sup_Mat} for details).

\emph{Observability in digital quantum computer}s. We expect the emergent
non-Hermitian COMs at EPs identified in this work can be observed
experimentally. The non-Hermitian Hamiltonian (\ref{eq: Heisenberg_nhh})
can be realized, for instance, by employing the weak measurement combined
with post-selections. More specifically, in a short period $dt$,
the evolution operator can be first trotterized into the unitary part
$\hat{U}(dt)\sim\exp\left[-idt\left(\hat{H}_{\text{XXX}}+\sum_{j=1}^{L}\hat{\sigma}_{j}^{x}\right)\right]$
and the nonunitary part $\exp\left(dt(1-\Delta)\sum_{j=1}^{L}\hat{\sigma}_{j}^{y}\right)$.
For experiments employing digital quantum computers, the unitary part
$\hat{U}(dt)$ can be realized by standard one- and two-bit gates.
The nonunitary part $\exp\left(dt(1-\Delta)\sum_{j=1}^{L}\hat{\sigma}_{j}^{y}\right)$
can be realized by coupling an ancilla qubit to the system's qubits
through controlled-NOT gates, performing projection measurement on
the ancilla qubit, and then postselecting the quantum state of the
complete system according to the measurement outcomes \citep{brun2008PRA}.
A concrete protocol for realizing a two-site non-Hermitian spin-$1/2$
Heisenberg chain with digital quantum computers is shown in Fig.~\ref{fig: IBMQC}(a)
(see Supplemental Material \cite{Sup_Mat} for more details), with
$\hat{U}_{\mathrm{I}}$ being a two-bit unitary operation that initializes
the state of the system, $\hat{R}\equiv(e^{i\pi/4}\hat{I}+e^{-i\pi/4}\sum_{a=x,y,z}\hat{\sigma}^{a})/2$
being a single qubit unitary operation acting on any qubit of the
system, and $\hat{U}_{\mathrm{I}}^{\mathrm{A}}\equiv e^{-i(\pi/4-(1-\Delta)dt)\hat{\sigma}^{y}}$
being a one-bit operation that initializes the state of the ancilla
qubit. The measurements on the ancilla qubit are performed along the
$z$-direction, i.e., with the corresponding set of measurement operators
being $\{|0\rangle\langle0|,|1\rangle\langle1|\}$. And trajectories
with the measurement outcome of the ancilla qubit being in $|1\rangle$
are rejected afterward.

As a direct verification, we directly simulate the dynamics of $\hat{C}_{1}$
for a two-site system using the protocol shown in Fig.~\ref{fig: IBMQC}(a)
on the IBM quantum circuit simulator, the \emph{qasm\_simulator},
with the noise from \emph{ibm\_lagos} \citep{cross2018ibm}. As shown
in Fig.~\ref{fig: IBMQC}(b), the expectation value of $\hat{C}_{1}$
evolves away from its initial value for finite $\Delta$, whereas
it keeps unchanged for $\Delta=0$, indicating the emergence of a
non-Hermitian COM at EP.

\emph{Conclusions}. We reveal a new class of distinct phenomena that
is hinged on exceptional points in the non-Hermitian quantum systems,
namely, the emergence of a series of non-Hermitian conservation laws.
These emergent conservation laws can be systematically identified
by employing generalized eigenvectors, and can be traced back to the
presence of emergent symmetries in the auxiliary \emph{Hermitian}
systems. Concrete simulations show that these emergent conserved dynamics
can be observed in current digital quantum computing systems. We believe
that our work will stimulate theoretical research on distinct properties
associated with EPs in interacting non-Hermitian quantum many-body
systems, particularly the dynamical behavior on a large time scale,
as well as experimental efforts in directly observing these non-Hermitian
conserved dynamics. Moreover, recent studies \cite{Marko_2019_PRL,Ilievski_2021_PRX,Ye_2022_PRL,David_2022_science}
have shown that conserved charges associated with the $\text{SU}(2)$
Lie group in the Hermitian Heisenberg chain can exhibit Kardar-Parisi-Zhang
(KPZ) dynamics in finite-temperature transport. This motivates the
intriguing possibility that similar KPZ dynamics may emerge for non-Hermitian
COMs at EPs, such as $\hat{C}_{2}$, which can be traced back to a
conserved charge associated with the $\text{SU}(2)$ Lie group in
the Hermitian Heisenberg chain. Another intriguing direction for future
research is the study of the emergence of conservation laws in quantum
systems subjected to microscopic uncertainties, such as quantum many-body
systems exposed to thermal noise and quantum circuits influenced by
random measurements, among other scenarios. 
\begin{acknowledgments}
This work is supported by the NKRDPC (Grant No.~2022YFA1405304),
NSFC (Grant No.~12275089), and the Guangdong Basic and Applied Basic
Research Foundation (Grants No.~2023A1515012800), and Guangdong Provincial
Key Laboratory (Grant No.~2020B1212060066). We acknowledge the use
of IBM Quantum services for this work. The views expressed are those
of the authors, and do not reflect the official policy or position
of IBM or the IBM Quantum team. 
\end{acknowledgments}

%

\clearpage{} 

\onecolumngrid 

\vspace{\columnsep} 
\begin{center}
{\large\textbf{Supplemental Material for ``Emergent non-Hermitian
conservation laws at exceptional points''}}{\large\par}
\par\end{center}

\vspace{\columnsep}


\twocolumngrid

\setcounter{equation}{0}

\setcounter{figure}{0}

\setcounter{page}{1}

\setcounter{section}{0}

\global\long\def\theequation{S\arabic{equation}}%
\global\long\def\thesection{S\arabic{section}}%
\global\long\def\thefigure{S\arabic{figure}}%
%
%

\section{COMs at EP constructed by generalized Eigenvectors}

Here, we derive the COMs associated with an EP of order $N$ in a
non-Hermitian system described by Hamiltonian $\hat{H}$. We consider
the generalized eigenvectors of the conjugated Hamiltonian $\hat{H}^{\dagger}$,
denoted as $|V_{1}\rangle,\dots,|V_{N}\rangle$, satisfying $\hat{H}^{\dagger}|V_{n}\rangle=E|V_{n}\rangle+|V_{n-1}\rangle$
for $1\leq n\leq N$ and $|V_{0}\rangle\equiv0$ . Any operators $\hat{O}$
within the subspace spanned by $\{|V_{n}\rangle|n=1,\dots,N\}$ can
be decomposed into the linear combinations of $\{\hat{V}_{i,j}\equiv|V_{i}\rangle\langle V_{j}||i,j=1,2,\dots,N\}$,
i.e., 
\begin{equation}
\hat{O}=\sum_{i,j}c_{i,j}\hat{V}_{i,j}.
\end{equation}
The COMs of $\hat{H}$ satisfy 
\begin{align}
0 & =\hat{H}^{\dagger}\hat{O}-\hat{O}\hat{H}\nonumber \\
 & =\sum_{i,j}c_{i,j}\hat{H}^{\dagger}\hat{V}_{i,j}-\sum_{i,j}c_{i,j}\hat{V}_{i,j}\hat{H}.\label{eq: deriving_COM_1}
\end{align}
By definition, we have the properties 
\begin{align}
\hat{H}^{\dagger}\hat{V}_{i,j} & =E\hat{V}_{i,j}+\hat{V}_{i-1,j},\\
\hat{V}_{i,j}\hat{H} & =E\hat{V}_{i,j}+\hat{V}_{i,j-1}.
\end{align}
Plugging in these formula into Eq.~(\ref{eq: deriving_COM_1}) we
obtain 
\begin{align}
0= & \sum_{i=1}^{L}\sum_{j=1}^{L}c_{i,j}\hat{V}_{i-1,j}-\sum_{i=1}^{L}\sum_{j=1}^{L}c_{i,j}\hat{V}_{i,j-1}\nonumber \\
= & \sum_{i=1}^{L-1}\sum_{j=1}^{L}c_{i+1,j}\hat{V}_{i,j}-\sum_{i=1}^{L}\sum_{j=1}^{L-1}c_{i,j+1}\hat{V}_{i,j}\nonumber \\
= & \sum_{i=1}^{L-1}\sum_{j=1}^{L-1}\left(c_{i+1,j}-c_{i,j+1}\right)\hat{V}_{i,j}\nonumber \\
 & +\sum_{i=1}^{L-1}\left(c_{i+1,L}\hat{V}_{i,L}-c_{L,i+1}\hat{V}_{L,i}\right).
\end{align}
The solutions are given by 
\begin{align}
c_{i+1,j}=c_{i,j+1} & \text{ for }i,j=1,\dots L-1,\\
c_{i+1,L}=c_{L,i+1}=0 & \text{ for }i=1,\dots L-1.
\end{align}
Especially, the non-vanishing coefficients give rise to $N$ linearly
independent COMs in Eq.~(4).

\section{Emergent symmetry in an auxiliary Hermitian system}

Here we demonstrate the emergent larger symmetry in an auxiliary Hermitian
system that corresponds to a generic non-Hermitian system at EP. It
will be convenient to choose appropriate basis vectors to represent
operators into matrices. More specifically, we can cast the Hamiltonian
matrix at EP into its Jordan normal form, given by $\oplus_{i=1}^{i_{\max}}(E_{i}\mathbf{I}_{N_{i}}+\mathbf{J}_{N_{i}}^{+})$,
where the $N_{i}\times N_{i}$ matrices $\mathbf{I}_{N_{i}}$ and
$\mathbf{J}_{N_{i}}^{+}$ are given by $(\mathbf{I}_{N_{i}})_{i,j}=\delta_{i,j}$
and $(\mathbf{J}_{N_{i}}^{+})_{i,j}=\delta_{i+1,j}$ with $\delta_{i,j}$
being the Kronecker delta function. Begin with the Jordan normal form,
we can convert all the $1$'s in $\mathbf{J}_{N_{i}}^{+}$ to non-zero
values by similarity transformations. This allows us to express an
arbitrary Hamiltonian matrix at the EP in the following way 
\begin{equation}
\mathbf{H}=\oplus_{i=1}^{i_{\max}}(E_{i}\mathbf{I}_{N_{i}}+\mathbf{S}_{N_{i}}^{x}+i\mathbf{S}_{N_{i}}^{y}),\label{eq: Jordan_Prime_Form_EP-1}
\end{equation}
where $\mathbf{S}_{N_{i}}^{x(y)(z)}$ is the $N_{i}\times N_{i}$
spin matrices for spin-$(N_{i}-1)/2$ operators along the $x(y)(z)$-direction
in the $z$-basis (the conventions $\hbar=1$ and $\mathbf{S}_{N_{i}=1}^{x}=\mathbf{S}_{N_{i}=1}^{y}=\mathbf{S}_{N_{i}=1}^{z}=0$
are adopted), and $E_{i}$ is the corresponding real eigenvalue for
the $i$th block of $\mathbf{H}$. Subsequently, we introduce a perturbation
to the Hamiltonian matrix (\ref{eq: Jordan_Prime_Form_EP-1}) by adding
a term, $-\oplus_{i=1}^{i_{\max}}i\Delta\mathbf{S}_{N_{i}}^{y}$,
\begin{equation}
\mathbf{H}(\Delta)=\stackrel[i=1]{i_{\max}}{\oplus}(E_{i}\mathbf{I}_{N_{i}}+\mathbf{S}_{N_{i}}^{x}+i(1-\Delta)\mathbf{S}_{N_{i}}^{y})\label{eq: Jordan_Prime_Perturbed-1}
\end{equation}
Under a specific perturbation with $2>\Delta>0$, we find that $\mathbf{S}_{N_{i}}^{x}+i(1-\Delta)\mathbf{S}_{N_{i}}^{y}$
can be transformed into a Hermitian matrix using the invertible matrix
$\mathbf{S}_{N_{i}}\equiv\exp(\mathbf{S}_{N_{i}}^{z}\ln\sqrt{\Delta/(2-\Delta)})$,
i.e., 
\begin{equation}
\mathbf{S}_{N_{i}}(\mathbf{S}_{N_{i}}^{x}+i(1-\Delta)\mathbf{S}_{N_{i}}^{y})\mathbf{S}_{N_{i}}^{-1}=\sqrt{\Delta(2-\Delta)}\mathbf{S}_{N_{i}}^{x}.
\end{equation}
Applying this transformation to all the blocks in $\mathbf{H}(\Delta)$
(using invertible matrix $\mathbf{S}\equiv\stackrel[i=1]{i_{\max}}{\oplus}\mathbf{S}_{N_{i}}$),
we obtain the Hermitian Hamiltonian matrix, $\mathbf{H}_{\text{A}}(\Delta)$,
i.e., 
\begin{align}
\mathbf{H}_{\text{A}}(\Delta) & =\mathbf{S}\mathbf{H}(\Delta)\mathbf{S}^{-1},\nonumber \\
 & =\stackrel[i=1]{i_{\max}}{\oplus}(E_{i}\mathbf{I}_{N_{i}}+\sqrt{\Delta(2-\Delta)}\mathbf{S}_{N_{i}}^{x}).\label{eq: NH2H-1}
\end{align}
Similar to the Hermitian counterpart of the non-Hermitian Heisenberg
chain, the Hermitian Hamiltonian $\mathbf{H}_{\text{A}}(\Delta)$
displays an exact symmetry generated by rotations along the $x$-direction,
as $\mathbf{S}_{N_{i}}^{x}$ represents the spin-$(N_{i}-1)/2$ operator
in this direction. As $\Delta$ approaches zero, the degeneracies
in $\mathbf{H}_{\text{A}}(\Delta\rightarrow0^{+})$ indicate the existence
of larger symmetries.

\section{Deriving the auxiliary Hermitian Hamiltonians}

In this section, we derive the auxiliary Hermitian Hamiltonians using
two approaches. The first approach involves solving the problem for
a small system and then generalizing the result to systems of arbitrary
size. The second approach utilizes the Jordan normal form and generalized
eigenvectors.

\subsection{The first approach: generalizing from a small system}

We begin by considering a single-spin system with the Hamiltonian
\begin{equation}
\hat{H}_{\text{ss}}=g\hat{\sigma}^{x}+ig(1-\Delta)\hat{\sigma}^{y}.
\end{equation}
Our goal is to find an invertible operator that maps $\hat{H}_{\text{ss}}$
to a Hermitian operator. A straightforward computation reveals that
the simplest such operator is $\hat{\mathcal{S}}_{\text{ss}}=\exp(\hat{\sigma}^{z}/2\ln\sqrt{\Delta/(2-\Delta)})$,
which describes a rotation around the $z$-axis with an imaginary
angle. Using this transformation, the Hermitian counterpart of the
Hamiltonian is given by 
\begin{equation}
\hat{\mathcal{S}}_{\text{ss}}\hat{H}_{\mathrm{ss}}\hat{\mathcal{S}}_{\text{ss}}^{-1}=g\sqrt{\Delta(2-\Delta)}\hat{\sigma}^{x}.
\end{equation}

Now, we extend this result to the many-body case, where the Hamiltonian
is given by: $\hat{H}_{\mathrm{NHS}}=\hat{H}_{\mathrm{XXX}}+g\sum_{j=1}^{L}\hat{\sigma}_{j}^{x}+ig(1-\Delta)\sum_{j=1}^{L}\hat{\sigma}_{j}^{y}$,
where $\hat{H}_{\text{XXX}}\equiv J\sum_{j=1}^{L-1}\sum_{a=x,y,z}\hat{\sigma}_{j}^{a}\hat{\sigma}_{j+1}^{a}$
is the Hamiltonian for the isotropic Heisenberg chain. Since $\hat{H}_{\text{XXX}}$
is already Hermitian, we seek a transformation operator that preserves
it while transforming the non-Hermitian terms. A natural generalization
of $\hat{\mathcal{S}}_{\text{ss}}$ that satisfies this requirement
is 
\begin{align}
\hat{\mathcal{S}} & =\exp(\sum_{j}^{L}\hat{\sigma}_{j}^{z}/2\ln\sqrt{\Delta/(2-\Delta)})\nonumber \\
 & =\otimes_{j=1}^{L}\exp(\hat{\sigma}_{j}^{z}/2\ln\sqrt{\Delta/(2-\Delta)}).
\end{align}
Remarkably, this transformation successfully maps $\hat{H}_{\mathrm{NHS}}$
to the following auxiliary Hermitian Hamiltonian 
\begin{equation}
\hat{\mathcal{S}}\hat{H}_{\mathrm{NHS}}\hat{\mathcal{S}}^{-1}=\hat{H}_{\mathrm{XXX}}+\sum_{j=1}^{L}\sqrt{\Delta(2-\Delta)}\hat{\sigma}_{j}^{x}.
\end{equation}
This demonstrates that the auxiliary Hermitian Hamiltonian can be
systematically obtained by constructing an appropriate similarity
transformation that generalizes from the single-spin case to the many-body
system.

\subsection{The second approach: using the Jordan normal form}

First, recall that the isotropic Heisenberg chain possesses $\text{SU}(2)$
symmetry. Specifically, this implies that $\hat{H}_{\text{XXX}}\equiv4J\sum_{j=1}^{L-1}\sum_{a=x,y,z}\hat{S}_{j}^{a}\hat{S}_{j+1}^{a}$
commutes with $\hat{S}^{2}=(\hat{S}^{x})^{2}+(\hat{S}^{y})^{2}+(\hat{S}^{z})^{2}$
and $\hat{S}^{z}$, where $\hat{S}^{x(y)(z)}\equiv\sum_{j=1}^{L}\hat{S}_{j}^{x(y)(z)}$.
We can label the energy eigenstates by $\{|n,s_{n},m_{s}\rangle\}$
($m_{s}=-s_{n},\ldots,s_{n}$), satisfying 
\begin{align}
\hat{H}_{\text{XXX}}|n,s_{n},m_{s}\rangle & =2gE_{n}|n,s_{n},m_{s}\rangle,\\
\hat{S}^{2}|n,s_{n},m_{s}\rangle & =(s_{n}+1)s_{n}|n,s_{n},m_{s}\rangle,\\
\hat{S}^{z}|n,s_{n},m_{s}\rangle & =m_{s}|n,s_{n},m_{s}\rangle,\\
\hat{S}^{\pm}|n,s_{n},m_{s}\rangle & =\sqrt{(s_{n}\mp m_{s})(s_{n}\pm m_{s}+1)}|n,s_{n},m_{s}\pm1\rangle.
\end{align}
Here, we have introduced an additional factor of $2g$ in the eigenvalues
$2gE_{n}$ for later convenience. Using the energy basis $\{|n,s_{n},m_{s}\rangle\}$,
we express the non-Hermitian Hamiltonian $\left(2g\right)^{-1}\hat{H}_{\mathrm{NHS}}=\left(2g\right)^{-1}\hat{H}_{\mathrm{XXX}}+\hat{S}^{x}+i(1-\Delta)\hat{S}^{y}$
in matrix form, with matrix elements 
\begin{align}
 & \langle n^{\prime},s_{n^{\prime}}^{\prime},m_{s^{\prime}}^{\prime}|\frac{1}{2g}\hat{H}_{\mathrm{NHS}}|n,s_{n},m_{s}\rangle\nonumber \\
= & E_{n}\delta_{n,n^{\prime}}\delta_{s_{n},s_{n^{\prime}}^{\prime}}\delta_{m_{s},m_{s^{\prime}}^{\prime}}+\langle n^{\prime},s_{n}^{\prime},m_{s}^{\prime}|\frac{\hat{S}^{+}+\hat{S}^{-}}{2}|n,s_{n},m_{s}\rangle\nonumber \\
 & +(1-\Delta)\langle n^{\prime},s_{n}^{\prime},m_{s}^{\prime}|\frac{\hat{S}^{+}-\hat{S}^{-}}{2}|n,s_{n},m_{s}\rangle\nonumber \\
= & E_{n}\delta_{n,n^{\prime}}\delta_{s_{n},s_{n^{\prime}}^{\prime}}\delta_{m_{s},m_{s^{\prime}}^{\prime}}\nonumber \\
 & +\delta_{n,n^{\prime}}\delta_{s_{n},s_{n^{\prime}}^{\prime}}\sqrt{s_{n}^{2}+s_{n}-m_{s}m_{s^{\prime}}^{\prime}}\frac{\delta_{m_{s}+1,m_{s^{\prime}}^{\prime}}+\delta_{m_{s},m_{s^{\prime}}^{\prime}+1}}{2}\nonumber \\
 & +(1-\Delta)\delta_{n,n^{\prime}}\delta_{s_{n},s_{n^{\prime}}^{\prime}}\sqrt{s_{n}^{2}+s_{n}-m_{s}m_{s^{\prime}}^{\prime}}\frac{\delta_{m_{s}+1,m_{s^{\prime}}^{\prime}}-\delta_{m_{s},m_{s^{\prime}}^{\prime}+1}}{2}.\label{eq:non-Hermit_mat_element}
\end{align}
We observe that the Hamiltonian is block-diagonalized.

The $(2s+1)\times(2s+1)$ spin matrices for $\mathbf{S}_{2s+1}^{x(y)}$
are defined by 
\begin{align}
\left(\mathbf{S}_{2s+1}^{x}\right)_{ab} & =\frac{\delta_{a,b+1}+\delta_{a+1,b}}{2}\sqrt{(s+1)(a+b-1)-ab},\\
i\left(\mathbf{S}_{2s+1}^{y}\right)_{ab} & =\frac{\delta_{a+1,b}-\delta_{a,b+1}}{2}\sqrt{(s+1)(a+b-1)-ab},
\end{align}
where $1\leq a,b\leq2s+1$. By reindexing $a,b$ as $m_{s}=a-s-1$
and $m_{s}^{\prime}=b-s-1$, we have 
\begin{align}
\left(\mathbf{S}_{2s+1}^{x}\right)_{m_{s},m_{s}^{\prime}} & =\frac{\delta_{m_{s},m_{s}^{\prime}+1}+\delta_{m_{s}+1,m_{s}^{\prime}}}{2}\sqrt{s^{2}+s-m_{s}m_{s}^{\prime}},\\
i\left(\mathbf{S}_{2s+1}^{y}\right)_{m_{s},m_{s}^{\prime}} & =\frac{\delta_{m_{s}+1,m_{s}^{\prime}}-\delta_{m_{s},m_{s}^{\prime}+1}}{2}\sqrt{s^{2}+s-m_{s}m_{s}^{\prime}}.
\end{align}
Substituting $\mathbf{S}_{2s+1}^{x}$ and $i\mathbf{S}_{2s+1}^{y}$
into Eq.~(\ref{eq:non-Hermit_mat_element}) and fixing $(n^{\prime},s_{n^{\prime}}^{\prime})=(n,s_{n})$,
we find 
\begin{align}
 & \langle n,s_{n},m_{s}^{\prime}|\frac{1}{2g}\hat{H}_{\mathrm{NHS}}|n,s_{n},m_{s}\rangle\nonumber \\
= & E_{n}\delta_{m_{s},m_{s}^{\prime}}+\left(\mathbf{S}_{2s_{n}+1}^{x}\right)_{m_{s},m_{s}^{\prime}}+i(1-\Delta)\left(\mathbf{S}_{2s_{n}+1}^{y}\right)_{m_{s},m_{s}^{\prime}}.
\end{align}
This reveals that, in the basis $\{|n,s_{n},m_{s}\rangle\}$, the
non-Hermitian Hamiltonian $\left(2g\right)^{-1}\hat{H}_{\mathrm{NHS}}$
indeed takes the form of Eq.~(\ref{eq: Jordan_Prime_Perturbed-1}).
Consequently, the auxiliary Hermitian Hamiltonian of $\left(2g\right)^{-1}\hat{H}_{\mathrm{NHS}}$
and the similarity transformation matrix in this basis are given by
\begin{align}
\mathbf{S} & =\oplus_{n}\exp(\mathbf{S}_{2s_{n}+1}^{z}\ln\sqrt{\Delta/(2-\Delta)}),\\
\mathbf{H}_{\text{A}}(\Delta) & =\oplus_{n}(E_{n}\mathbf{I}_{2s_{n}+1}+\sqrt{\Delta(2-\Delta)}\mathbf{S}_{2s_{n}+1}^{x}).
\end{align}
By rewriting the matrix expression in terms of operators and multiplying
by $2g$, we recover the Hermitian counterpart of $\hat{H}_{\mathrm{NHS}}$.
This leads us to the expressions presented in the main text 
\begin{align}
\hat{\mathcal{S}} & =\exp(\hat{S}^{z}\ln\sqrt{\Delta/(2-\Delta)})\nonumber \\
 & =\otimes_{j=1}^{L}\exp(\hat{S}_{j}^{z}\ln\sqrt{\Delta/(2-\Delta)}),\\
\hat{H}_{\mathrm{AHS}} & =\hat{H}_{\mathrm{XXX}}+2g\sqrt{\Delta(2-\Delta)}\hat{S}^{x}.
\end{align}
Thus, we have successfully derived the auxiliary Hermitian Hamiltonian
using the Jordan normal form approach.

\subsection{Application of the second approach in other systems}

Remarkably, the derivations above rely solely on the presence of an
$\text{SU}(2)$-symmetric Hamiltonian, $\hat{H}_{\text{XXX}}$, and
a set of operators, $\hat{S}^{x(y)(z)}$, that realize the $\text{SU}(2)$
group generators. As a result, this derivation can be readily extended
to any system featuring an $\text{SU}(2)$-symmetric Hamiltonian and
its corresponding generators.

A key example is the isotropic Heisenberg chain with local transverse
fields, described by \cite{Wang2023arXiv}: 
\begin{equation}
\hat{H}_{\mathrm{XXX}}+g\sum_{j\in\mathcal{M}}\hat{\sigma}_{j}^{x}+ig(1-\Delta)\sum_{j\in\mathcal{M}}\hat{\sigma}_{j}^{y},
\end{equation}
where $\mathcal{M}$ denotes the set of local sites subjected to the
transverse fields. Following the same reasoning as in the above derivation,
we find that its Hermitian counterpart is simply given by: $\hat{H}_{\mathrm{XXX}}+\sum_{j\in\mathcal{M}}\sqrt{\Delta(2-\Delta)}\hat{\sigma}_{j}^{x}$.

Another example is the Fermi-Hubbard model with a non-Hermitian term,
given by \cite{Wang2023arXiv}: 
\begin{align}
\hat{H}_{\mathrm{NHF}}= & \hat{H}_{\mathrm{FHM}}+g\sum_{j=1}^{L}(\hat{c}_{j,\uparrow}^{\dagger}\hat{c}_{j,\downarrow}+\hat{c}_{j,\downarrow}^{\dagger}\hat{c}_{j,\uparrow})\nonumber \\
 & +ig(1-\Delta)\sum_{j=1}^{L}(\hat{n}_{j,\uparrow}-\hat{n}_{j,\downarrow}),
\end{align}
where $\hat{H}_{\mathrm{FHM}}\equiv-J\sum_{j=1}^{L-1}\sum_{\sigma}(\hat{c}_{j,\sigma}^{\dagger}\hat{c}_{j+1,\sigma}+\mathrm{h}.\mathrm{c}.)+U\sum_{j=1}^{L}\hat{n}_{j,\uparrow}\hat{n}_{j,\downarrow}$
is the Hermitian Fermi-Hubbard model. In this case, $\hat{H}_{\mathrm{FHM}}$
exhibits $\text{SU}(2)$ symmetry, with the associated generators
defined as:

\begin{align}
\hat{S}_{\text{FHM}}^{x} & =\frac{1}{2}\sum_{j=1}^{L}(\hat{c}_{j,\uparrow}^{\dagger}\hat{c}_{j,\downarrow}+\hat{c}_{j,\downarrow}^{\dagger}\hat{c}_{j,\uparrow}),\\
\hat{S}_{\text{FHM}}^{y} & =\frac{i}{2}\sum_{j=1}^{L}(-\hat{c}_{j,\uparrow}^{\dagger}\hat{c}_{j,\downarrow}+\hat{c}_{j,\downarrow}^{\dagger}\hat{c}_{j,\uparrow}),\\
\hat{S}_{\text{FHM}}^{z} & =\frac{1}{2}\sum_{j=1}^{L}(\hat{n}_{j,\uparrow}-\hat{n}_{j,\downarrow}).
\end{align}
These operators satisfy the standard $\mathfrak{su}(2)$ Lie algebra:
\begin{align}
[\hat{S}_{\text{FHM}}^{x},\hat{S}_{\text{FHM}}^{y}] & =i\hat{S}_{\text{FHM}}^{z},\\{}
[\hat{S}_{\text{FHM}}^{z},\hat{S}_{\text{FHM}}^{x}] & =i\hat{S}_{\text{FHM}}^{y},\\{}
[\hat{S}_{\text{FHM}}^{y},\hat{S}_{\text{FHM}}^{z}] & =i\hat{S}_{\text{FHM}}^{x}.
\end{align}
Expressing the non-Hermitian Hamiltonian in terms of these generators,
we obtain: 
\begin{equation}
\hat{H}_{\mathrm{NHF}}=\hat{H}_{\mathrm{FHM}}+2g\hat{S}_{\text{FHM}}^{x}+i2g(1-\Delta)\hat{S}_{\text{FHM}}^{z}.
\end{equation}
Following the same reasoning as in the approaches discussed above,
we identify the similarity transformation and the corresponding Hermitian
counterpart: 
\begin{align}
\hat{\mathcal{S}} & =\exp(\hat{S}_{\text{FHM}}^{y}\ln\sqrt{(2-\Delta)/\Delta}),\\
\hat{\mathcal{S}}\hat{H}_{\mathrm{NHF}}\hat{\mathcal{S}}^{-1} & =\hat{H}_{\mathrm{FHM}}+2g\sqrt{\Delta(2-\Delta)}\hat{S}_{\text{FHM}}^{x}.
\end{align}

These results demonstrate the general applicability of our approach
to non-Hermitian Hamiltonians consisting of a Hermitian $\text{SU}(2)$-symmetric
component and a non-Hermitian term constructed from $\text{SU}(2)$
group generators.

\section{Deriving COMs at EP from auxiliary Hermitian system}

The correspondence relation in Eq.~(8) can be generalized to systems
described in (\ref{eq: Jordan_Prime_Form_EP-1}) and (\ref{eq: NH2H-1}).
Denoting $N$ linearly independent COMs at EP within a relevant Jordan
block of size $N$ as $\{\mathbf{C}_{n}\}$, we can find their one-to-one
correspondence to $N$ linearly independent COMs, $\{\mathbf{C}_{n}^{\text{A}}\}$,
of auxiliary Hermitian system, 
\begin{equation}
\mathbf{C}_{n}=\lim_{\Delta\rightarrow0^{+}}\mathbf{S}_{N}^{\dagger}\mathbf{C}_{n}^{\text{A}}\mathbf{S}_{N},\quad n=1,\dots,N.\label{eq: Correspondence_CA2CEP-1}
\end{equation}
Following the logic when deriving Eq.~(8), i.e., first identify $\{\mathbf{C}_{n}\}$
and then figure out $\{\mathbf{C}_{n}^{\text{A}}\}$, Eq.~(\ref{eq: Correspondence_CA2CEP-1})
can be easily established. However, in this way, we still don't know
whether there is a correspondence at the EP for those COMs ($\{\mathbf{C}^{\text{A}}\}-\{\mathbf{C}_{n}^{\text{A}}\}$)
of auxiliary Hermitian system that are not in the set $\{\mathbf{C}_{n}^{\text{A}}\}$.
Here, we shall show that beginning with all the COMs of auxiliary
Hermitian system, $\{\mathbf{C}^{\text{A}}\}$, we can find that only
a subset $\{\mathbf{C}_{n}^{\text{A}}\}\subsetneq\{\mathbf{C}^{\text{A}}\}$
can give rise to the COMs at EP, $\{\mathbf{C}_{n}\}$, via (\ref{eq: Correspondence_CA2CEP-1}).

We focus on a single Jordan block, where the Hamiltonian matrix of
dimension $N$ is $E\mathbf{I}_{N}+\mathbf{S}_{N}^{x}+i(1-\Delta)\mathbf{S}_{N}^{y}$
with its Hermitian counterpart in the region $2>\Delta>0$ being $E\mathbf{I}_{N}+\sqrt{\Delta(2-\Delta)}\mathbf{S}_{N}^{x}$.
At $\Delta=0$, the Hermitian Hamiltonian becomes an identity matrix.
Therefore, a given COM, $\mathbf{C}^{\text{A}}$, in the complete
set, $\{\mathbf{C}^{\text{A}}\}$, can be expressed by finite-valued
elements $(\mathbf{C}^{\text{A}})_{a,b}=c_{a,b}$.

First, we assess $\mathbf{S}_{N}^{\dagger}\mathbf{C}^{\text{A}}\mathbf{S}_{N}$
as $\Delta\rightarrow0^{+}$. The transformation matrix is given by
\begin{align}
\mathbf{S}_{N}^{\dagger}=\mathbf{S}_{N}= & e^{\mathbf{S}_{N}^{z}\ln\sqrt{\frac{\Delta}{2-\Delta}}}\nonumber \\
= & \left(\begin{array}{cccc}
\delta^{(\frac{N-1}{2})/2}\\
 & \delta^{(\frac{N-3}{2})/2}\\
 &  & \ddots\\
 &  &  & \delta^{(\frac{1-N}{2})/2}
\end{array}\right),
\end{align}
where $\delta\equiv\Delta/(2-\Delta)$. Both $\Delta$ and $\delta$
are infinitesimals, with $\delta^{k}$ representing an infinitesimal
of the $k$th order \citep{bradley2010cauchy}. The counterpart of
$\mathbf{C}^{\text{A}}$ is then given by 
\begin{equation}
\mathbf{S}_{N}^{\dagger}\mathbf{C}^{\text{A}}\mathbf{S}_{N}=\left(\begin{array}{cccc}
c_{1,1}\delta^{\frac{N-1}{2}} & c_{1,2}\delta^{\frac{N-2}{2}} & \cdots & c_{1,N}\\
c_{2,1}\delta^{\frac{N-2}{2}} & c_{2,2}\delta^{\frac{N-3}{2}} & \cdots & c_{2,N}\delta^{-\frac{1}{2}}\\
\vdots & \vdots & \ddots & \vdots\\
c_{N,1} & c_{N,2}\delta^{-\frac{1}{2}} & \cdots & c_{N,N}\delta^{-\frac{N-1}{2}}
\end{array}\right).\label{eq: SCS}
\end{equation}
Before continuing, let's see what the typical COM, $\mathbf{I}_{N}\in\{\mathbf{C}^{\text{A}}\}$,
of auxiliary Hermitian system corresponds to near the EP. Here, Eq.~(\ref{eq: SCS})
becomes 
\begin{equation}
\mathbf{S}_{N}^{\dagger}\mathbf{I}_{N}\mathbf{S}_{N}=\left(\begin{array}{cccc}
\delta^{\frac{N-1}{2}} & 0 & \cdots & 0\\
0 & \ddots & \vdots & \vdots\\
\vdots & \cdots & \delta^{-\frac{N-3}{2}} & 0\\
0 & \cdots & 0 & \delta^{-\frac{N-1}{2}}
\end{array}\right).
\end{equation}
We notice that as $\delta\rightarrow0^{+}$, the last two diagonal
elements are $\delta^{-\frac{N-1}{2}}$ and $\delta^{-\frac{N-3}{2}}$,
which are divergences of different orders if $N>3$. Note that the
physical observables in a \emph{finite-dimensional} Hilbert space
are represented by matrices with finite norms. Therefore, for $N>3$,
$\text{I}_{N}$ has no corresponding observable in the non-Hermitian
system at EP. Only a subset of $\{\mathbf{C}^{\text{A}}\}$ can have
meaningful correspondence at EP, and the condition satisfied by this
subset is 
\begin{equation}
\max(\{c_{a,b}\})\sim\begin{cases}
\delta^{0}, & N+1\geq a+b\geq2\\
\delta^{\frac{a+b-N-1}{2}}, & 2N\geq a+b\geq N+2
\end{cases}.\label{eq: coeff_bounded_constrain}
\end{equation}

Next, we consider the matrix $\mathbf{C}^{\prime}$ ($\{\mathbf{C}^{\prime}\}\subsetneq\{\mathbf{C}^{\text{A}}\}$)
constituted by those elements satisfying (\ref{eq: coeff_bounded_constrain}).
From this, we determine that the expression for $[E\mathbf{I}_{N}+\sqrt{\Delta(2-\Delta)}\mathbf{S}_{N}^{x},\mathbf{C}^{\prime}]$
is represented by the matrix entries 
\begin{align}
 & \left([\mathbf{S}_{N}^{x},\mathbf{C}^{\prime}]\right)_{a,b}\nonumber \\
= & \frac{\sqrt{(N+1-a)(a-1)}}{2}c_{a-1,b}-\frac{\sqrt{(N+1-b)(b-1)}}{2}c_{a,b-1}\nonumber \\
 & +\frac{\sqrt{(N-a)a}}{2}c_{a+1,b}-\frac{\sqrt{(N-b)b}}{2}c_{a,b+1},
\end{align}
where $c_{0,b}=c_{N+1,b}=c_{a,0}=c_{a,N+1}=0$. Applying $\mathbf{S}_{N}^{\dagger}$
and $\mathbf{S}_{N}$ to both sides of $(2-\Delta)\delta^{1/2}[\mathbf{S}_{N}^{x},\mathbf{C}^{\prime}]$,
we obtain 
\begin{align}
 & \mathbf{S}_{N}^{\dagger}[E\mathbf{I}_{N}+\sqrt{\Delta(2-\Delta)}\mathbf{S}_{N}^{x},\mathbf{C}^{\prime}]\mathbf{S}_{N}\nonumber \\
= & (2-\Delta)\delta^{\frac{1}{2}}\mathbf{S}_{N}^{\dagger}[\mathbf{S}_{N}^{x},\mathbf{C}^{\prime}]\mathbf{S}_{N}\nonumber \\
= & \left(\begin{array}{cccc}
\bullet_{1,1}\delta^{\frac{N-1}{2}} & \bullet_{1,2}\delta^{\frac{N-2}{2}} & \cdots & \bullet_{1,N}\\
\bullet_{2,1}\delta^{\frac{N-2}{2}} & \bullet_{2,2}\delta^{\frac{N-3}{2}} & \cdots & \bullet_{2,N}\delta^{-\frac{1}{2}}\\
\vdots & \vdots & \ddots & \vdots\\
\bullet_{N,1} & \bullet_{N,2}\delta^{-\frac{1}{2}} & \cdots & \bullet_{N,N}\delta^{-\frac{N-1}{2}}
\end{array}\right),\label{eq: RHS_correspondence}
\end{align}
where $\bullet_{a,b}\equiv(2-\Delta)\delta^{1/2}\left([\mathbf{S}_{N}^{x},\mathbf{C}^{\prime}]\right)_{a,b}$.
We want to find those $\mathbf{C}^{\prime}$ making Eq.~(\ref{eq: RHS_correspondence})
to be $\mathbf{0}$ in the $\Delta\rightarrow0^{+}$ limit, since
this is equivalent to 
\begin{align}
 & \left[E\mathbf{I}_{N}+\mathbf{S}_{N}^{x}+i(1-\Delta)\mathbf{S}_{N}^{y}\right]^{\dagger}\left(\mathbf{S}_{N}^{\dagger}\mathbf{C}^{\prime}\mathbf{S}_{N}\right)\nonumber \\
= & \left(\mathbf{S}_{N}^{\dagger}\mathbf{C}^{\prime}\mathbf{S}_{N}\right)\left[E\mathbf{I}_{N}+\mathbf{S}_{N}^{x}+i(1-\Delta)\mathbf{S}_{N}^{y}\right].
\end{align}

For $N+1\geq a+b\geq2$, $\{c_{a-1,b},c_{a,b-1}\}$ are at most of
order $\delta^{0}$, and $\{c_{a+1,b},c_{a,b+1}\}$ are likewise at
most of order $\delta^{0}$. Consequently, these $(a,b)$ entries
are physical zeros for any $c_{a,b}$ satisfying (\ref{eq: coeff_bounded_constrain}).
The remaining matrix in (\ref{eq: RHS_correspondence}) is 
\begin{equation}
\left(\begin{array}{cccc}
0 & 0 & \cdots & 0\\
\vdots & \vdots & \iddots & \bullet_{2,N}\delta{}^{-\frac{1}{2}}\\
\vdots & 0 & \iddots & \vdots\\
0 & \bullet_{N,2}\delta^{-\frac{1}{2}} & \cdots & \bullet_{N,N}\delta^{-\frac{N-1}{2}}
\end{array}\right).
\end{equation}

For $2N\geq a+b\geq N+2$, $\{c_{a-1,b},c_{a,b-1}\}$ are at most
of order $\delta^{(a+b-N-2)/2}$, and $\{c_{a+1,b},c_{a,b+1}\}$ are
at most of order $\delta^{(a+b-N)/2}$. Hence, the $(a,b)$ entry
of the above matrix is 
\begin{align}
 & \bullet_{a,b}\delta^{-\frac{a+b-N-1}{2}}\nonumber \\
= & (2-\Delta)\delta^{\frac{1}{2}}\left([S_{N}^{x},C^{\prime}]\right)_{a,b}\delta^{-\frac{a+b-N-1}{2}}\nonumber \\
= & (2-\Delta)\delta^{-\frac{a+b-N-2}{2}}\frac{\sqrt{(N+1-a)(a-1)}}{2}c_{a-1,b}\nonumber \\
 & -(2-\Delta)\delta^{-\frac{a+b-N-2}{2}}-\frac{\sqrt{(N+1-b)(b-1)}}{2}c_{a,b-1}\nonumber \\
 & +(2-\Delta)\delta^{-\frac{a+b-N-2}{2}}\frac{\sqrt{(N-a)a}}{2}c_{a+1,b}\nonumber \\
 & -(2-\Delta)\delta^{-\frac{a+b-N-2}{2}}\frac{\sqrt{(N-b)b}}{2}c_{a,b+1}.
\end{align}
Notice that $\{\delta^{-(a+b-N-2)/2}c_{a+1,b},\delta^{-(a+b-N-2)/2}c_{a,b+1}\}$
are maximally of order $\delta^{1}$, and $\{\delta^{-(a+b-N-2)/2}c_{a-1,b},\delta^{-(a+b-N-2)/2}c_{a,b-1}\}$
are maximally of order $\delta^{0}$. The only way to ensure $\bullet_{a,b}\delta^{-(a+b-N-1)/2}=0$
is 
\begin{equation}
\frac{c_{a-1,b}}{c_{a,b-1}}=\frac{\sqrt{(N+1-b)(b-1)}}{\sqrt{(N+1-a)(a-1)}}.
\end{equation}
We obtain $N-1$ sets of the above equations since $2N\geq a+b\geq N+2$.
These $N-1$ independent solutions, together with $c_{N,N}$ maximally
taking values of order $\delta^{(N-1)/2}$, yield $N$ linearly independent
COMs, represented as 
\begin{equation}
\{\mathbf{C}_{n}^{\text{A}}\}\subsetneq\{\mathbf{C}^{\prime}\}\subsetneq\{\mathbf{C}^{\text{A}}\}.
\end{equation}
These directly give rise to $N$ linearly independent COMs in the
non-Hermitian system at the EP, i.e., $\mathbf{C}_{n}=\mathbf{S}_{N}^{\dagger}\mathbf{C}_{n}^{\text{A}}\mathbf{S}_{N}$
for $n=1,\dots,N$.

In summary, the constraints in (\ref{eq: coeff_bounded_constrain})
(lead to $\{\mathbf{C}^{\prime}\}$) and the vanishing matrix entries
in (\ref{eq: RHS_correspondence}) (lead to $\{\mathbf{C}_{n}^{\text{A}}\}$)
yield $N$ linearly independent COMs at the EP, $\{\mathbf{C}_{n}\}$.

\section{Conserved dynamics in the presence of noise}

\begin{figure}
\includegraphics[height=1.8in]{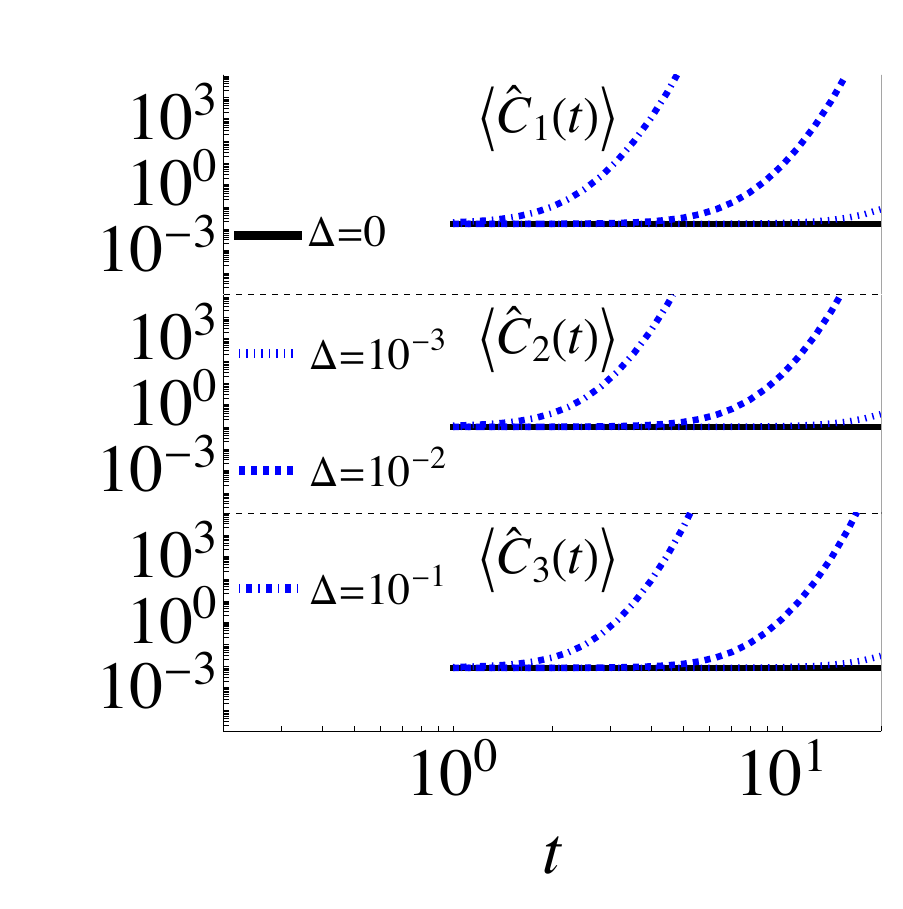}\caption{Time evolution of $\langle\hat{C}_{1,\,2,\,3}\rangle(t)$ averaged
over $500$ stochastic realizations of $\{\xi_{j}\}$, both at and
away from the EP, for the modified non-Hermitian Heisenberg chain
with noise $\hat{H}_{\text{noise}}[\xi]$. The system parameters are
$L=6$, $J=1$, and the initial state is $|\psi(t=0)\rangle=2^{-L}\sum_{\{\sigma_{j}^{z}\}}|\sigma_{1}^{z}\rangle\otimes\cdots\otimes|\sigma_{L}^{z}\rangle$.}\label{fig: C_1_2_3_time_evolution_Inhomo}
\end{figure}

Here, we consider a modified model in which the complex Zeeman term
in Eq.~(1) includes site-dependent random noise. The Hamiltonian
of this model, $\hat{H}_{\text{noise}}[\xi]$, is given by 
\begin{equation}
\hat{H}_{\text{noise}}[\xi]=\hat{H}_{\mathrm{XXX}}+g\sum_{j=1}^{L}\hat{\sigma}_{j}^{x}+ig\sum_{j=1}^{L}(1-\Delta\xi_{j})\hat{\sigma}_{j}^{y},
\end{equation}
where $\xi_{j}$ is a real random number drawn from a Gaussian distribution
with zero mean and unit variance, i.e., $\langle\xi_{j}\rangle=0$
and $\langle\xi_{i}\xi_{j}\rangle=\delta_{ij}$. The time evolution
of the first three conserved quantities $\langle\hat{C}_{1,\,2,\,3}\rangle(t)$
is shown in Fig.~\ref{fig: C_1_2_3_time_evolution_Inhomo}. As is
evident from the figure, the conservation laws approximately hold
true over a finite timescale when the parameter $\Delta$ is nonzero
but small.

\section{Experimental Protocol with quantum circuit}

Here, we demonstrate how the quantum circuit in Fig.~3(a) can simulate
the dynamics of the two-site non-Hermitian spin-$1/2$ Heisenberg
chain with Hamiltonian $\hat{H}_{\mathrm{NHS}}$.

Within the first time period $dt$, the unitary operation $\hat{U}_{\text{I}}$
prepares the initial state $|\psi(t=0)\rangle=2^{-2}\sum_{\sigma_{1}^{z},\sigma_{2}^{z}}|\sigma_{1}^{z}\rangle\otimes|\sigma_{2}^{z}\rangle$
from $|\uparrow\rangle\otimes|\uparrow\rangle$. (The $|0\rangle$
state in Fig.~3(a) represents $|\uparrow\rangle$ satisfying $\hat{\sigma}^{z}|\uparrow\rangle=|\uparrow\rangle$.)
Subsequently, the unitary part of the dynamics is conducted by $\hat{U}(dt)$,
resulting in the intermediate state spanned by eigenstates of $\hat{\sigma}^{y}$,
$\sum_{\sigma_{1}^{y},\sigma_{2}^{y}}c_{\sigma_{1}^{y},\sigma_{2}^{y}}|\sigma_{1}^{y}\rangle\otimes|\sigma_{2}^{y}\rangle$.
In the small $dt$ limit, the non-unitary part can be decomposed as
\begin{align}
 & \exp\left(\tilde{dt}\sum_{j=1}^{L}\hat{\sigma}_{j}^{y}\right)\nonumber \\
\sim & \left(\frac{\cos\tilde{dt}}{\sqrt{2}}\right)^{-L}\prod_{j=1}^{L}\left(\frac{\cos\tilde{dt}}{\sqrt{2}}+\frac{\sin\tilde{dt}}{\sqrt{2}}\hat{P}_{j,\uparrow_{y}}-\frac{\sin\tilde{dt}}{\sqrt{2}}\hat{P}_{j,\downarrow_{y}}\right),
\end{align}
where $\tilde{dt}\equiv(1-\Delta)dt$, $\hat{P}_{j,\uparrow_{y}}\equiv\frac{1+\hat{\sigma}_{j}^{y}}{2}$,
$\hat{P}_{j,\downarrow_{y}}\equiv\frac{1-\hat{\sigma}_{j}^{y}}{2}$.
The $\left(\frac{\cos\tilde{dt}}{\sqrt{2}}+\frac{\sin\tilde{dt}}{\sqrt{2}}\hat{P}_{1,\uparrow_{y}}-\frac{\sin\tilde{dt}}{\sqrt{2}}\hat{P}_{1,\downarrow_{y}}\right)$
is implemented as follows. First, the ancilla qubit is reset to state
\begin{align}
 & e^{-i(\pi/4-\tilde{dt})\hat{\sigma}^{y}}|0\rangle_{a}\nonumber \\
= & \frac{1}{\sqrt{2}}\left(\cos\tilde{dt}+\sin\tilde{dt}\right)|0\rangle_{a}+\frac{1}{\sqrt{2}}\left(\cos\tilde{dt}-\sin\tilde{dt}\right)|1\rangle_{a}.
\end{align}
Second, perform unitary rotations $\hat{R}=e^{i\frac{\pi}{4}}\frac{\hat{I}}{2}+e^{-i\frac{\pi}{4}}\frac{\sum_{a=x,y,z}\hat{\sigma}^{a}}{2}$
satisfying $\hat{R}|\uparrow_{y}\rangle(|\downarrow_{y}\rangle)=|\uparrow\rangle(|\downarrow\rangle)$
on the first qubit, resulting in state $\sum_{\sigma_{1}^{y},\sigma_{2}^{y}}c_{\sigma_{1}^{y},\sigma_{2}^{y}}\left(\hat{R}|\sigma_{1}^{y}\rangle\right)\otimes|\sigma_{2}^{z}\rangle$.
Third, couple the first qubit with the ancilla qubit by a controlled-NOT
gate ($|0\rangle\langle0|\otimes\hat{I}_{a}+|1\rangle\langle1|\otimes\hat{\sigma}_{a}^{x}$)
so that the resulting state is 
\begin{align}
 & \sum_{\sigma_{2}^{y}}c_{\uparrow_{y},\sigma_{2}^{y}}\left(\hat{R}|\uparrow_{y}\rangle\right)\otimes|\sigma_{2}^{z}\rangle\otimes\left(e^{-i(\pi/4-\tilde{dt})\hat{\sigma}^{y}}|0\rangle_{a}\right)\nonumber \\
+ & \sum_{\sigma_{2}^{y}}c_{\downarrow_{y},\sigma_{2}^{y}}\left(\hat{R}|\downarrow_{y}\rangle\right)\otimes|\sigma_{2}^{z}\rangle\otimes\left(\hat{\sigma}_{a}^{x}e^{-i(\pi/4-\tilde{dt})\hat{\sigma}^{y}}|0\rangle_{a}\right).
\end{align}
Fourth, measure the ancilla qubit and discard those shots with measured
$|1\rangle_{a}$, so that the remaining state is 
\begin{align}
 & \sum_{\sigma_{2}^{y}}c_{\uparrow_{y},\sigma_{2}^{y}}\left(\hat{R}|\uparrow_{y}\rangle\right)\otimes|\sigma_{2}^{z}\rangle\otimes\frac{\cos\tilde{dt}+\sin\tilde{dt}}{\sqrt{2}}|0\rangle_{a}\nonumber \\
+ & \sum_{\sigma_{2}^{y}}c_{\downarrow_{y},\sigma_{2}^{y}}\left(\hat{R}|\downarrow_{y}\rangle\right)\otimes|\sigma_{2}^{z}\rangle\otimes\frac{\cos\tilde{dt}-\sin\tilde{dt}}{\sqrt{2}}|0\rangle_{a}.
\end{align}
Fifth, perform the inverse rotation $\hat{R}^{-1}$ on the first qubit
and drop the ancilla qubit state, the result coincides with acting
$\left(\frac{\cos\tilde{dt}}{\sqrt{2}}+\frac{\sin\tilde{dt}}{\sqrt{2}}\hat{P}_{1,\uparrow_{y}}-\frac{\sin\tilde{dt}}{\sqrt{2}}\hat{P}_{1,\downarrow_{y}}\right)$
on the left of $\sum_{\sigma_{1}^{y},\sigma_{2}^{y}}c_{\sigma_{1}^{y},\sigma_{2}^{y}}|\sigma_{1}^{y}\rangle\otimes|\sigma_{2}^{z}\rangle$.
We can perform the above protocol for the second qubit in the same
way so that the system's final state with two measured $|0\rangle_{a}$
is given by 
\begin{equation}
\left(\frac{\cos\left[(1-\Delta)dt\right]}{\sqrt{2}}\right)^{L}\exp\left(dt\hat{H}_{\mathrm{NHS}}\right)|\psi(t=0)\rangle.
\end{equation}

\end{document}